\documentclass{elsarticle}
\usepackage{graphicx}
\usepackage{bm}
\usepackage{enumerate}
\usepackage{amssymb,amsmath}
\begin{document}
\title{Numerical micromagnetism of strong inhomogeneities}
\author{Christian Andreas}  
\address{Peter Gr\"unberg Institut (PGI-6),
Forschungszentrum   J\"ulich    GmbH,   D-52428   J\"ulich,   Germany}
\address{Institut de Physique et Chimie des Mat\'eriaux de Strasbourg, Universit\'e de Strasbourg, CNRS UMR 7504, Strasbourg, France} 
\author{Sebastian Gliga}
\address{Laboratory for Micro- and Nanotechnology, Paul Scherrer Institute, 5232 Villigen PSI, Switzerland}
\address{Laboratory for Mesoscopic Systems, Department of Materials, ETH Zurich, 8093 Zurich, Switzerland}
\author{Riccardo Hertel\fnref{cor1}}
\ead{hertel@ipcms.unistra.fr}
\address{Institut de Physique et Chimie des Mat\'eriaux de Strasbourg, Universit\'e de Strasbourg, CNRS UMR 7504, Strasbourg, France}
\fntext[cor1]{Corresponding author: Riccardo Hertel, Tel.: +33 38810 7263; Fax: +33 38810 7249  }
\date{\today}
\begin{abstract}
The size of micromagnetic structures, such as domain walls or vortices, is comparable to the exchange length of the ferromagnet. Both, the exchange length of the stray field $l_s$ and the magnetocrystalline exchange length $l_k$, are material-dependent quantities that usually lie in the nanometer range. This emphasizes the theoretical challenges associated with the mesoscopic nature of micromagnetism: the magnetic structures are much larger than the atomic lattice constant, but at the same time much smaller than the sample size. In computer simulations, the smallest exchange length serves as an estimate for the largest cell size admissible to prevent appreciable discretization errors. This general rule is not valid in special situations where the magnetization becomes particularly inhomogeneous. When such strongly inhomogeneous structures develop, micromagnetic simulations inevitably contain systematic and numerical errors. It is suggested to combine micromagnetic theory with a Heisenberg model to resolve such problems. We analyze cases where strongly inhomogeneous structures pose limits to standard micromagnetic simulations, arising from fundamental aspects as well as from numerical drawbacks.\end{abstract}
 \maketitle
\section{Introduction}
The theory of micromagnetism was established firmly more than fifty years ago \cite{brown_micromagnetics_1963,brown_magnetostatic_1962,kronmuller_general_2007,aharoni_introduction_2000,hubert_magnetic_2012}. At that time, analytic calculations were performed according to the specific problem that was studied, {\em e.g.}, a one-dimensional magnetic domain wall \cite{bloch_zur_1932,neel_energie_1955}, a chain of vortices and antivortices in a cross-tie domain wall \cite{middelhoek_domain_1963}, or a Bloch point singularity \cite{doring_point_1968,feldtkeller_mikromagnetisch_1965}. When studying these problems, the mathematical methods, the approximations, and the limits of the model were clearly described. They represented the starting point of the specific study. Spectacular progress in computational micromagnetism has led to a shift in this domain of solid state theory. Micromagnetic simulations are now a commonly used, convenient, reliable and important means of 
research to investigate the properties of ferromagnetic nanostructures. While this has helped considerably in disseminating the fundamentals of micromagnetism to a broad scientific community, the limits of the range of validity of the theory have received less attention.
As a result, many powerful and easy-to-use programs are sometimes applied to problems that go beyond the range of validity of the underlying theory. 
A typical situation where this occurs is when strongly inhomogeneous magnetic structures develop. This article aims to point out some pitfalls that one encounters in numerical micromagnetics in the case of highly inhomogeneous structures and discusses possibilities of treating such pathological cases.

The article is structured as follows. After a description of the basic equations and the fundamental assumptions on which the theory of micromagnetism is based, the difference between discretization errors and methodological errors is briefly discussed in section \ref{err-types-sec}.  The importance of the exchange lengths as the typical length scale above which static micromagnetic theory is valid is recalled in section \ref{wall-sect} using the example of a one-dimensional domain wall. The limiting case where the domain wall width collapses to zero is discussed in section \ref{BP-sect}, which describes Bloch point singularities and their behavior in numerical studies. Using the exchange length as a reference scale, in section \ref{analysis-sect} we analyze, compare and quantify analytic and numeric errors that occur when a spin spiral has a periodicity smaller than the exchange length, thereby connecting the cases of a smooth and an abrupt transition. Finally, in section \ref{fast-sect}, a short discussion of the limits of validity of the dynamic equations in the case of ultrafast processes emphasizes that strongly inhomogeneous structures are only one of several cases where micromagnetism reaches its limits.

\section{Basic equations\label{basic-sec}}
Calculating the spatial and temporal evolution of the magnetization in a ferromagnet is the central task of micromagnetic theory. The magnetization $\bm{M}(\bm{r},t)$ is defined as the density of magnetic moments \cite{kronmuller_general_2007},

\begin{equation}\label{conti}
\bm{M}(\bm{r},t)=\frac{1}{V(\bm{r})}\sum\limits_{{\bm R}\in V(\bm{r})}\bm{\mu_R}(t)
\end{equation}
where $V(\bm{r})$ is the volume of a mesoscopic part of the sample around $\bm{r}$, and $\bm{\mu_R}$ is a microscopic magnetic moment at the point $\bm{R}$ within $V(\bm{r})$. Even though the microscopic magnetic moment at a point within a ferromagnet results from the electronic structure of the material, it can be assumed for simplicity that $\bm{\mu_R}$ are atomistic magnetic moments. The volume $V(\bm{r})$ is sufficiently small so that the magnetization is homogeneous and hence the magnitude of the magnetization $M_s=|\bm{M}(\bm{r},t)|$ is a constant material parameter: the spontaneous magnetization.

The transition from a set of $i$ microscopic magnetic moments located at atomic lattice sites $\bm{R}$, {\em i.e.}, $\left\{\bm{\mu_R}^i\right\}$ to a continuous vector field of the magnetization $\bm{M}(\bm{r},t)$ is also performed for the calculation of the magnetostatic field. The dipolar magnetic field $\bm{h}_{\rm dip}$ at the position $\bm{R}$ results from the sum of the field of each magnetic moment,
\begin{equation}
\bm{h}_{{\rm dip},\bm{R}}= \frac{1}{4\pi} \sum\limits_{\bm{R'}\neq\bm{R}}\left\{\frac{3\left[ \bm{\mu_{R'}}(t)\cdot(\bm{R}-\bm{R'})\right]\cdot(\bm{R}-\bm{R'})}{\left|\bm{R}-\bm{R'}\right|^5}
-\frac{\bm{\mu_{R'}}(t)}{\left|\bm{R}-\bm{R'}\right|^3} \right\}
\end{equation}
can be converted into an integral over the density of magnetic moments, leading to 

\begin{equation}\label{gradU}
\bm{H}_{\rm dip}=-\bm{\nabla}U(\bm{r},t)
\end{equation}
where the magnetostatic potential $U(\bm{r},t)$ is

\begin{equation}\label{potential}
U(\bm{r},t)=-\iiint\limits_{\rm volume}\frac{{\rm div}\bm{M}(\bm{r'},t)}{4\pi|\bm{r}-\bm{r'}|}{\rm d}^3r'+ \oint\limits_{{\rm surface}}\frac{\bm{M}(\bm{r'},t)\cdot\hat{\bm{n}}(\bm{r'})}{4\pi|\bm{r}-\bm{r'}|}\,{\rm d}S'\, .
\end{equation}


The first integral on the right hand side runs over the volume of the ferromagnet, and the second integral over the surface \cite{brown_magnetostatic_1962}. Here $\hat{\bm{n}}$ is the outward oriented unit vector perpendicular to the surface and $ {\rm d}S$ is an infinitesimal surface element. 

Micromagnetic theory contains a further energy term which is based on the transition from a set of discrete magnetic moments to a continuous density of magnetic moments: the exchange interaction. In its simplest form, the interatomic exchange is given by 
\begin{equation}\label{hei}
E_{\rm xc}=-J_{ij}\sum\limits_{NN,i\neq j}\hat{\bm{\mu}}_i\cdot\hat{\bm{\mu}}_j=E_0+J_{ij}\sum\limits_{NN}\hat{\bm{\mu}}_i(\hat{\bm{\mu}}_i-\hat{\bm{\mu}}_j)
\end{equation}
where $E_{\rm xc}$ is the exchange energy, ``NN" denotes a summation over nearest neighbors, $\hat{\bm{\mu}}_i$ is the normalized magnetic moment $\hat{\bm{\mu}}_i=\bm{\mu}_i/|\bm{\mu}_i|$, and $J_{ij}$ is the Heisenberg exchange integral. In Eq.~(\ref{hei}), $E_0$ is an energetic offset that can be omitted; thereby obtaining a positive-definite expression for $E_{\rm xc}$.
Assuming a small misalignment between neighboring magnetic moments, {\em i.e.},
\begin{equation}\label{small-ang}
|\hat{\bm{\mu}}_i-\hat{\bm{\mu}}_j|\ll 1,
\end{equation}
a Taylor expansion of this term leads to the micromagnetic form of the exchange interaction
\begin{equation}\label{exc-c}
e_{\rm xc}=A\left[\left(\frac{\partial\bm{m}}{\partial x}\right)^2+\left(\frac{\partial\bm{m}}{\partial y}\right)^2+\left(\frac{\partial\bm{m}}{\partial z}\right)^2\right]
\end{equation}
where $A$ is the exchange constant \cite{kittel_physical_1949} and $\bm{m}(\bm{r},t)=\bm{M}(\bm{r},t)/M_s$ is the reduced magnetization.

The equations $(\ref{conti}$), (\ref{gradU}), ($\ref{potential}$), and $(\ref{exc-c})$ provide a framework in which the ferromagnet is treated as a continuum, without considering explicitly atomistic effects of the ferromagnet. In micromagnetic theory, the impact of the crystalline structure on the magnetic properties is accounted for by means of material parameters, which depend on the position. 
An example is the magneto-crystalline anisotropy, which in its simplest form is uniaxial and is described by an energy density term 
\begin{equation}
e_{\rm ani}(\bm{r},t)=K\left\{1-\left[\hat{\bm{k}}\cdot\bm{m}(\bm{r},t)\right]^2\right\}
\end{equation}
where $K$ is the uniaxial anisotropy constant and $\hat{\bm{k}}$ is a unit vector parallel to the easy axis.

The equation of motion of the vector field of the magnetization $\bm{M}(\bm{r},t)$ is the Landau-Lifshitz-Gilbert equation \cite{landau_theory_1935,gilbert_phenomenological_2004}
\begin{equation}\label{LLG}
\frac{{\rm d}}{{\rm d}t}\bm{M}(\bm{r},t)=-\gamma \bm{M}(\bm{r},t)\times\bm{H}_{{\rm eff}}({\bm r},t)+\alpha\frac{\bm{M}(\bm{r},t)}{M_s}\times\frac{{\rm d}}{{\rm d}t}\bm{M}(\bm{r},t)\,.
\end{equation}
where $\alpha$ is the Gilbert damping parameter and $\gamma$ is the gyromagnetic ratio.
The effective field $\bm{H}_{\rm eff}$ contains contributions of the external field, the internal magnetosotatic field, the exchange field and the anisotropy field. 

\section{Types of errors in the calculation of strongly inhomogeneous structures\label{err-types-sec}}
In spite of spectacular advances in numerical methods, including powerful GPU-accelerated simulations \cite{kakay_speedup_2010,li_graphics_2010} and the near-perfect reproduction of experimental observations \cite{hertel_three-dimensional_2005,cherifi_virgin_2005}, it remains inevitable that results from micromagnetic codes contain two types of errors. The first is the discretization error; the second error arises from the limitations of the underlying theory. Both errors are usually negligible, but can become significant in cases where the structure of the magnetization becomes exceptionally inhomogeneous.

Discretization errors stem from the use of a finite set of data points. The data provided by simulation results at these discretization points constitutes an approximation of the magnetization field $\bm{M}(\bm{r},t)$, which is assumed to be continuous, {\em i.e.}, smooth and with a well-defined value at any point. Based on the computed set of data, an approximation for the field can be achieved, {\em e.g.}, by a piecewise constant or a piecewise linear representation of $\bm{M}$ in the region between the discretization points. The quality of this approximation obviously depends on the density of discretization points, and the discretization errors go to zero in the limit of infinitely small cell sizes. While this limit cannot be achieved in practice, it is often possible to extrapolate the data to zero cell size by using a set of different grids.

The second error arises from the micromagnetic theory itself. One of its fundamental assumptions is the condition (\ref{small-ang}), which states that the magnetization changes slowly on the length scale of the atomistic lattice parameter. This is justified by the dominant influence of the ferromagnetic exchange on short length scales, which tends to align neighboring spins parallel to each other. The assumption of small angles between neighboring spins is usually a very good approximation, but for a correct interpretation of simulation data it is important to identify the exceptions in which this assumption fails. 
 
In the following we shall illustrate and quantify the impact of these two types of errors in the case of strongly inhomogeneous magnetic structures. A rigorous, quantitative analysis of numerical errors would generally require an important arsenal of methodologies which are specific to the applied numerical method. Our approach will be more heuristic, but it should capture the essential features in sufficient detail. It will consist in studying situations where analytic results are available, and comparing them with the computed data.

\section{Exchange lengths and 180$^\circ$ domain walls\label{wall-sect}}

\begin{figure}
\begin{center}
\includegraphics[width=7cm]{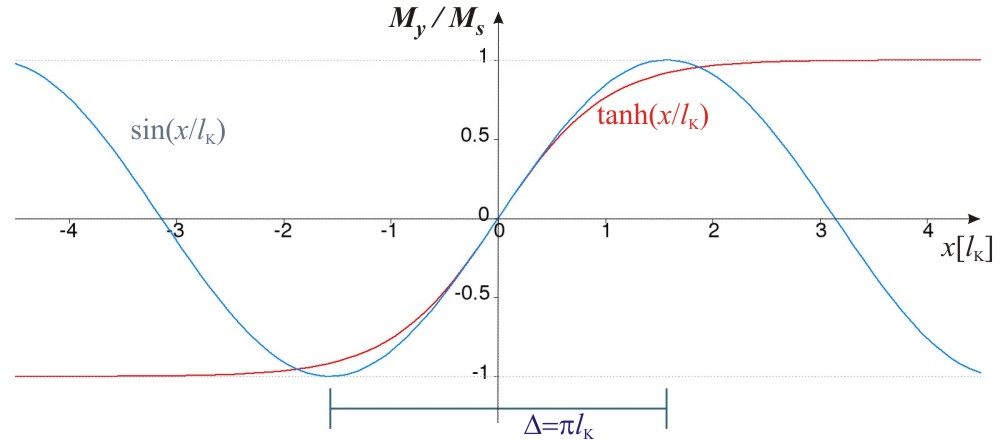}
\end{center}
\caption{\label{tnh} The red line displays the profile of an idealized one-dimensional 180$^\circ$ domain wall in an infinitely extended ferromagnet with uniaxial anisotropy. The direction of the easy axis is along $y$ and the red line displays the $y$-component of the normalized magnetization. In the case of a Bloch wall, the $x$ component is zero for all $x$. The corresponding domain wall width $\Delta=\pi l_K$ according to Lilley is sketched below. Close to the middle of the domain wall, $x=0$, the domain wall profile is very similar to that of a spin spiral with wave length $2\pi l_K=2\Delta$, displayed in blue.}
\end{figure}

Let us first recall the textbook example of a Bloch wall in a bulk ferromagnet with uniaxial anisotropy \cite{aharoni_introduction_2000,hubert_magnetic_2012,bloch_zur_1932,landau_theory_1935}, a case where the assumption of smooth changes of the magnetization is appropriate. The analytic solution for the profile of a $180^\circ$ domain wall is  obtained by minimizing a functional containing two energy terms. The one-dimensional variational problem then yields the typical kink shape of the domain wall profile shown in Fig.~\ref{tnh}. It displays the value of the $y$-component of the reduced magnetization $m_y(x)$ as a function of the position $x$. The easy axis points along $y$ and the domain wall centered at $x=0$. The boundary conditions are $\lim_{x\to\infty}m_y(x) =-1$ and $\lim_{x\to\infty}m_y(x) =+1$. Spherical variables are used to ensure that $M_s={\rm const.}$, and one obtains an angle $\phi=\phi(x)$ which defines the components of the magnetization $m_x=\cos\phi(x)$, $m_y=\sin\phi(x)$. It is assumed that the magnetization varies only along $x$, and that $m_z=0$. The latter condition ascertains that the rotation of the magnetization is perpendicular to the domain wall plane, as is the case for a Bloch wall. 

According to Lilley \cite{lilley_energies_1950} the domain wall width $\Delta=\pi l_k$ is proportional to the exchange length $l_K$. The one-dimensional Bloch wall profile may therefore serve to introduce the exchange length $l_K=(A/K)^{1/2}$ as a micromagnetic length scale, which can be considered as a tradeoff between the competing tendencies of the exchange interaction and the magneto-crystalline anisotropy.  With the given boundary conditions and the constraint $M_s={\rm const.}$, the magneto-crystalline anisotropy alone would favor a $180^\circ$ transition from $m_y=-1$ to $m_y=+1$ on a length scale as short as possible, while the exchange energy would be reduced by spreading the domain wall over a region as large as possible. In a similar fashion, N\'eel walls in thin films \cite{neel_energie_1955} lead to another exchange length $l_s=[2A/(\mu_0 M_s^2)]^{1/2}$, which results from the competition between the magnetostatic and the exchange interaction. 

For a given ferromagnetic material, $l=\min(l_K,l_s)$ describes the typical size of inhomogeneities, like domain walls or vortices. This role of the exchange lengths has been discussed in detail, {\em e.g.}, by Kronm\"uller \cite{kronmuller_mikromagnetische_1962}. In all practical cases $l$ is much larger than the atomic lattice constant, with typical values in the range between a few nm and several tens of nanometers. This provides {\em a posteriori} the justification for the approximation described by eq.~(\ref{small-ang}), according to which inhomogeneities of the magnetization are negligible on the atomic length scale. 

In view of this interpretation of $l$ as the characteristic size of inhomogeneous magnetic structures, it is clear that discretization cells in simulations should not be larger than $l$ in order to capture the details of the magnetic structure. This notwithstanding, there can be situations where finer grids will be required or coarser grids admissible. Since $l$ generally represents an estimate for the upper limit of the cell size, the question may arise whether a lower limit exists below which cell sizes are not allowed. The answer is no: cell sizes can be arbitrarily small; even smaller than the atomic lattice constant. 
This can be readily understood on the basis of the domain wall profile shown in Fig.~\ref{tnh}, where the components of the magnetization $M_x$, $M_y$, $M_z$ are defined within any arbitrarily small interval $[x, x+{\rm d}x]$. Likewise, the density of discretization points in simulation studies can be arbitrarily high. Choosing too small cell sizes is normally a waste of computational resources, but there is no fundamental criterion that poses an obstacle.

Having outlined the importance of exchange lengths as well as the role of the discretization cell size and the fundamental assumptions of micromagnetism, we can now investigate how systematic and numerical errors emerge in the case of strongly inhomogeneous structures, {\em i.e.}, when a significant inhomogeneity extends over a scale smaller than $l$. An inhomogeneity can be defined as significant if the magnetization changes its direction by at least 90$^\circ$. Practical cases of strongly inhomogeneous magnetic structures include the following situations:
\begin{itemize}
\item If the system is in a non-equilibrium state, {\em e.g.}, when a domain structure is suddenly exposed to a strong external field, the conditions used to derive $l$ do not apply, and the domain wall width can become significantly smaller. In this context, the velocity-dependent compression of field-driven domain walls and the resulting increase of exchange energy have been used by D\"oring to define the domain wall mass \cite{doring_uber_1948}.
\item The exchange length only plays a role if the ferromagnet can be considered as homogeneous, which is the case for or a single-crystalline material or for an amorphous alloy with sufficiently small phases. Granular composite magnets with strongly different anisotropies and exchange-coupled and phases can lead to magnetic inhomogeneities in the sub-nm range \cite{hrkac_role_2010, kronmuller_micromagnetism_1997}.
\item Topological defects represent singularities of the magnetization where $M_s$ collapses to zero at a point around which the magnetization becomes maximally inhomogeneous. Such Bloch points or Feldtkeller singularities \cite{doring_point_1968,feldtkeller_mikromagnetisch_1965,elias_magnetization_2011} can play a decisive role in magnetic switching processes \cite{arrott_point_1979,thiaville_micromagnetic_2003}. The assumption (\ref{small-ang}) does not hold in these cases, rendering such structures a significant problem for micromagnetic simulations. 
\end{itemize}

\section{Simulation of singularities\label{BP-sect}}

Singularities like Bloch points are not unusual in continuum theories. A well-known example is the laminar flow of an incompressible liquid around a sharp corner, like a $90^\circ$ edge in a vessel. This results in a divergence of the flow velocity near the edge, which is only calculated correctly if the discretization mesh is fine enough (see, {\em e.g.}, p.~456 of Ref.~\cite{zienkiewicz_finite_2005}). 

\begin{figure}
\begin{center}
\includegraphics[width=7cm]{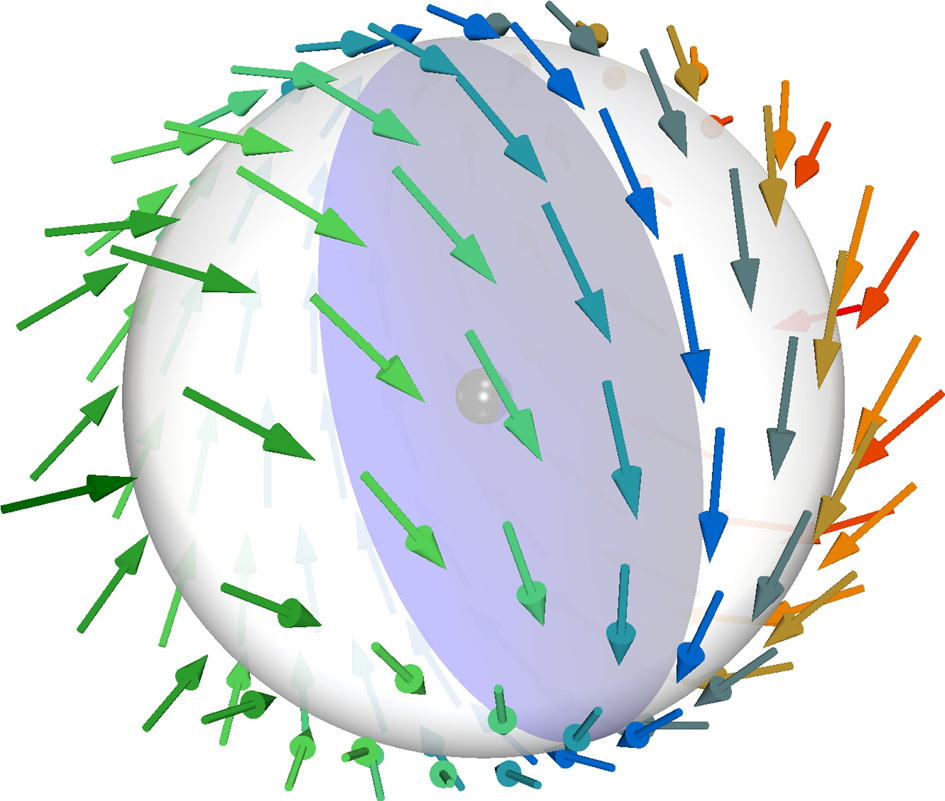}
\end{center}
\caption{\label{bp-sketch}The image displays an example of the micromagnetic structure in the vicinity of a Bloch point. Such point singularities in the three-dimensional vector field $\bm{M}(\bm{r},t)$ generate highly frustrated and strongly inhomogeneous regions. In the center of the Bloch point, no direction of the magnetization can be defined. Each subset of the magnetization field defined on the surface of a small sphere containing the Bloch point fills the entire directional space, 
{\em i.e.}, it contains at least one magnetization vector pointing in any chosen direction.}
\end{figure}

The usual way of dealing with such singularities in simulation studies (while remaining in the framework of the continuum theory) is an extrapolation of the computed values to infinite discretization density. The convergence rate and the limiting value can thereby be determined as the cells become smaller. 
In this context it is irrelevant whether the singularity is real in a physical sense. Obviously the velocity of the molecules in a liquid flowing around a corner does not really diverge, and the exchange energy density in a Bloch point is not infinite. 
The occurrence of such singularities indicates that something is wrong with the fundamental assumptions of the model: A real liquid is not completely incompressible, a corner is not perfectly sharp, and neighboring atomic magnetic moments may sometimes be significantly misaligned.
Nevertheless, if the continuum theory yields singularities, the computational task consists in providing solutions of these problematic regions as accurately as possible, while the scientific task consists in evaluating the validity of the theory and the underlying assumptions. It is noteworthy that such singularities in continuum theories never affect the system in a ``catastrophic" way, meaning that the overall result remains correct. In the case of a Bloch point, {\em e.g.}, the total energy is finite, in spite of the divergence of the exchange energy density. 
\begin{figure}
\begin{center}
\includegraphics[width=7cm]{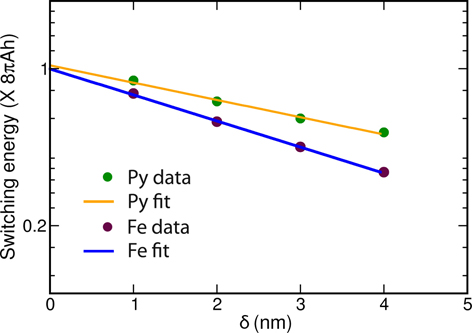}
\end{center}
\caption{\label{bp-nrg-fig}The energy required to switch a vortex core is equal to the energy needed to generate a Bloch point. According to analytic theory, the value of this energy is $8\pi Ah$ (see text). Numerical simulations systematically underestimate this switching energy, even if the cell size $\delta$ is well below the exchange length $l_K$. However, an extrapolation to zero cell size reconciles the results, yielding perfect agreement with analytic theory. This is confirmed for different magnetic materials (from Ref.\cite{gliga_energy_2011}).}
\end{figure}

An example of the method of extrapolating results obtained with different discretization densities is shown in Fig.~\ref{bp-nrg-fig}, where the threshold energy required to initiate a vortex core reversal is determined. Only by using this method, an almost perfect agreement with theoretical predictions could recently be obtained in micromagnetic simulation of vortex core switching processes. As discussed in Refs.~\cite{van_waeyenberge_magnetic_2006,hertel_ultrafast_2007} the switching of a vortex core in a thin-film element requires the formation of a Bloch point traversing the sample along the film thickness. The analytic value of the energy $E_{\rm BP}$  of a  Bloch point in a thin-film element of thickness $h$ has been calculated analytically, yielding \cite{polyakov_metastable_1975,tretiakov_vortices_2007}
\begin{equation}
E_{\rm BP}=8\pi Ah
\end{equation}
where $A$ is the exchange constant. The universality of this result is shown in Fig.~\ref{bp-nrg-fig}, where the switching energy was calculated for different materials. Details on the micromagnetic problem, the sample size and shape and the physical conclusions of this result are summarized in Ref.~\cite{gliga_energy_2011}. Note that without the extrapolation, the computed switching energy barrier is significantly underestimated -- even when the cell size is well below the exchange length. This is typical for singularities, where the convergence rate, {\em i.e.}, the amount by which the numeric error diminishes with increasing discretization density, is significantly lower than in the absence of singularities, cf., {\em e.g.}, Ref.~\cite{rave_magnetic_2000}.

\section{Non-singular strong inhomogeneities\label{analysis-sect}} 
The singular behavior discussed in the previous section and the smooth transition within a 180$^\circ$ wall (cf.~Fig.~\ref{tnh}) can be considered as two limiting cases: In the case of a Bloch point treated within the framework of micromagnetism, the magnetization changes its direction by 180$^\circ$ inside an infinitely small volume element ${\rm d}V$, while in the case of a Bloch {\em wall} the 180$^\circ$ transition occurs smoothly over a distance in the order of $\pi l_K$. The mesoscopic region between those limiting cases leads to numerical and fundamental difficulties that unfold as the inhomogeneity of a micromagnetic structure becomes more pronounced. These problems are exclusively connected to the exchange interaction, since it is the only micromagnetic term containing the assumption of small-angle interatomic variations of magnetic moments according to Eq.~(\ref{small-ang}).

\begin{figure}
\begin{center}
\includegraphics[width=7cm]{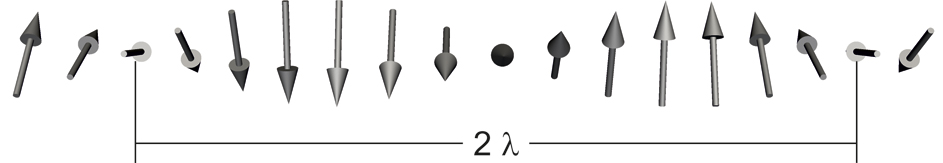}
\end{center}
\caption{\label{spiral}A simple spin spiral of periodicity $\lambda$ serves as a model system to study the behavior of numerical and systematic errors occurring when the exchange energy of an increasingly homogeneous structure are calculated. The direction in which the spin spiral is oriented is irrelevant, but it is the same for the entire three-dimensional sample. This does not represent an equilibrium arrangement. As was done, {\em e.g.}, in Ref.~\cite{hertel_finite_2002}, the structure is artificially imposed in order to monitor the differences between numerical and analytic results.}
\end{figure}
In order to simplify the analysis of gradually increasing inhomogeneities, we consider a spin spiral rather than the usual $\tanh(x)$ profile of the one-dimensional domain wall. As shown in Fig.~\ref{tnh}, a spin spiral of wave length $l_K$ is very similar to the profile of a domain wall of width $\Delta=\pi l_K$ near the center of the domain wall. The advantage of the spin spiral over the $\tanh(x)$ profile is a homogeneous exchange energy density, which allows to monitor deviations more easily than in the case of a Bloch wall profile. The spin spiral is defined by a periodicity $\lambda$ with the angle $\varphi(x)=x\pi/\lambda$, $m_x=\sin\varphi(x)$, $m_y=\cos\varphi(x)$, and the homogeneous exchange energy density of such a spin spiral has the value 
\begin{equation}\label{bp-nrg}
e_{\rm xc}^{\mu}(\lambda)=A\frac{\pi^2}{\lambda^2}\quad .
\end{equation}
The singularity for the case $\lambda\to 0$ is immediately recognized. In physical terms, the singularity is easy to understand as a result of the constraint of constant $M_s$ combined with the continuous definition of $\bm{M}(r)$ in space. While in a real magnet the lattice constant $a$ represents a lower bound for $\lambda$, the micromagnetic equations allow for arbitrarily small values of $\lambda$. Moreover, the highest-energy case of antiparallel magnetic moments in eq.~(\ref{hei}) yields a value that is lower than that of a spin spiral $\lambda=a$ calculated with the continuum term (\ref{exc-c}).

The analytic value according to eq.~(\ref{bp-nrg}) is derived from classic analytic theory and can serve as a reference to determine the accuracy of the numerically computed exchange energy density. For a detailed comparison of various effects emerging as $\lambda$ decreases, we compare three values of the energy density:
\begin{enumerate}
\item The analytic value $e_{\rm xc}^{\mu}(\lambda)$ is the exact solution of the micromagnetic equation (\ref{exc-c}),  which can be calculated irrespective of the validity of the approximations and assumptions on which micromagnetic theory is based.
\item The numerical value $e_{\rm xc}^{\rm FEM}(\lambda,\delta)$ is the value of the exchange energy density calculated with our code uning the finite-element method (FEM) for a given spin spiral of length $\lambda$ and a discretization size $\delta$.
\item The Heisenberg-term $e_{\rm xc}^{\rm Hei}(\lambda)$ is the value of the exchange energy density calculated according to eq.~(\ref{hei}). Although the value is calculated numerically, this term is not affected by discretization error since no interpolation scheme is used. This term is calculated for different crystalline structures, and in some cases we extend the interaction to include the contribution of atomic moments located within a shell much larger than the nearest-neighbor approximation.
\end{enumerate}

\begin{figure}\begin{center}
\includegraphics[width=7cm]{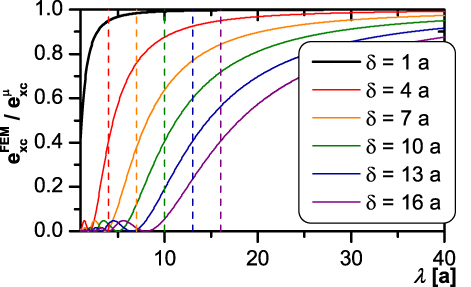}
\end{center}
\caption{\label{dscr-err}The ratio of the computed micromagnetic exchange energy $e^{\rm FEM}_{\rm xc}(\lambda,\delta)$ and the corresponding analytic value $e^\mu_{\rm xc}(\lambda)$ reaches very low values for small $\lambda$ and large discretization cell size 
$\delta$. The drop reflects the inability of the discretized solution to capture the drastic spike in the exchange energy density as $\lambda$ goes to zero. The dashed lines represent the discretization size $\delta$ in units of the lattice constant $a$, with the same colors as the $e^{\rm FEM}_{\rm xc}(\lambda,\delta)$ lines, respectively. The material is BCC n.n. \cite{material}}
\end{figure}

The evolution of these three energy terms as a function of $\lambda$ allows to discriminate between numerical and model-related errors and quantify the impact of these errors. We use our micromagnetic code {\tt TetraMag} to evaluate the exchange energy density for different cell sizes $\delta$ and discard the regions near the boundary to ensure that the results are not influenced by finite-size effects. In our case, the exchange field is calculated with standard FEM techniques, as described in Ref.~\cite{hertel_guided_2007}, but the numerical results and the behavior of discretization errors should not depend significantly on the numerical method that is used. The result is shown in Fig.~\ref{dscr-err}, where the ratio of the computed energy density $e_{\rm xc}^{\rm FEM}$ over $e_{\rm xc}^{\mu}$ is plotted for different cell sizes $\delta$. The length of the spiral $\lambda$ is given in units of the lattice atomic constant $a$. This should not distract from the fact that, so far, we have not left the territory of micromagnetic theory. The sharp drop of the curves near $\lambda=0$ shows that the numerical values cannot follow the steep increase of $e_{\rm exc}^{\mu}(\lambda)$ as $\lambda$ decreases. The numerically calculated value systematically underestimates the exchange energy. In usual micromagnetic structures, however, the error is small: If the domain wall width $\Delta$ extends over only 30 lattice constants, a cell size $\delta$ of about seven lattice constants will already lead to very good results. In practice, the values of $\Delta$ are even larger and the error smaller. 

\begin{figure}
\begin{center}
\includegraphics[width=7cm]{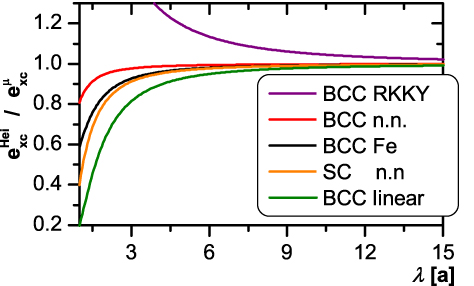}
\end{center}
\caption{\label{hei-mu}The micromagnetic approximation for the exchange energy $e^\mu_{\rm xc}$ according to eq.~(\ref{exc-c}) ceases to be valid for strongly inhomogeneous structures. The Heisenberg term, which can take different forms depending on the number of interacting atoms and the value of the exchange integral, assumes in any case a finite value, whereas the continuum expression diverges as $\lambda$ goes to zero. Well before $\lambda=a$ poses a natural limit for the largest inhomogeneity in the Heisenberg model, the equations yield significantly different results. 
The materials are chosen according to \cite{material}.}
\end{figure}

Next we analyze to which extent the analytic exchange energy density $e_{\rm xc}^{\mu}$, which is derived using a small-angle approximation and a transition to a continuum, deviates from an ``exact" model, where the exchange interaction is calculated with a classical Heisenberg-type summation as described in eq.~(\ref{hei}). The results are shown in Fig.~\ref{hei-mu}. The comparison is made here for different types of exact models in which a different number of interacting neighbors and different crystal structures have been taken into account.

For all but one of the cases, the energy density $e^\mu_{xc}(\lambda)$ calculated with the continuum expression (\ref{exc-c}) overestimates the exchange energy of strongly inhomogeneous structures. The only case that we found which deviates from this systematic tendency is one in which the assumption of strict ferromagnetic exchange is dropped, where an oscillatory RKKY-type exchange \cite{ruderman_indirect_1954,kasuya_theory_1956,yosida_magnetic_1957} is assumed, a large number of interacting atoms is considered. The resulting interaction favors an overall ferromagnetic order, but weakens it by an antiferromagnetic contribution.

\begin{figure}
\centerline{\includegraphics[width=7cm]{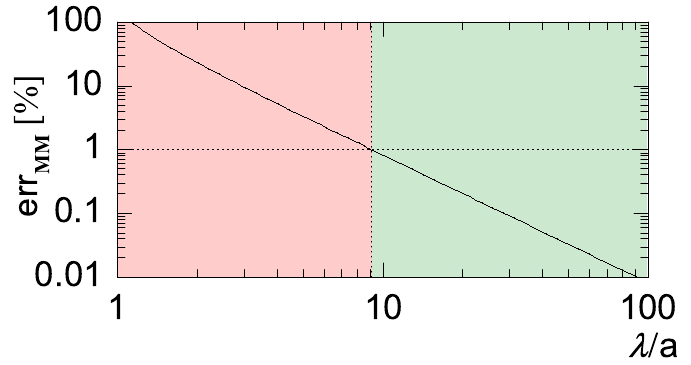}}
\caption{\label{e-hei-mu}Another way of representing the deviation shown in Fig.~\ref{hei-mu} is the error that is connected with the use of $e^\mu_{\rm xc}(\lambda)$ as a function of $\lambda$, expressed here in units of the lattice constant $a$. If $\lambda$ remains below about 15 unit cells, the error is smaller than $1\%$ and decreases rapidly with increasing $\lambda$. In the low-$\lambda$ regime, however, the errors due to the micromagnetic approximation become rapidly very significant.}
\end{figure}

The magnitude of the error connected with the use of the micromagnetic term for the exchange energy density according to eq.~(\ref{exc-c}) rather than the Heisenberg exchange interaction of eq.~(\ref{hei}) is shown in Fig.~\ref{e-hei-mu} as a function of the spin spiral length $\lambda$. In this case the energy density $e_{\rm xc}^{\rm Hei}$ is calculated with a Heisenberg-type summation over nearest-neighbors and compared with the analytic term $e^\mu_{\rm xc}(\lambda)$ derived from the micromagnetic approximation.

\begin{figure}
\centerline{\includegraphics[width=7cm]{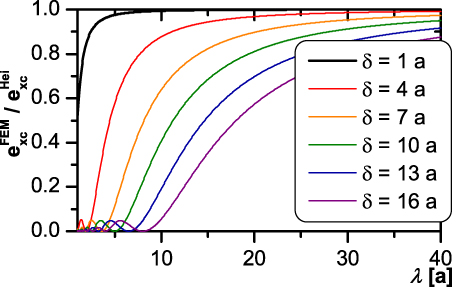}}
\caption{\label{e-fem_mu} The graph shows the numerical value $e^{\rm FEM}_{\rm xc}(\lambda,\delta)$ in units of the exchange energy calculated with the Heisenberg model as a function of $\lambda$. The behavior is very similar to that shown in Fig.~\ref{dscr-err}. This shows that, for all practical cases, the discretization error dominates largely over the errors resulting from the use of eq.~(\ref{exc-c}). In spite of their systematically opposite tendency, the two types of errors do not cancel each other. The material is BCC n.n. \cite{material}.}
\end{figure}

As a result it can be stated that in the framework of numerical micromagnetism, highly inhomogeneous structures result in both, numerical errors and methodological errors. These errors systematically have opposite sign, which at first may give hope that the results of numerical micromagnetic simulations could remain reliable if the different types of errors compensated each other; at least to a good extent. This is not the case, as shown in Fig.~\ref{e-fem_mu}, which displays the computed value $e_{\rm xc}^{\rm FEM}$ divided by the Heisenberg term $e_{\rm xc}^{\rm Hei}$. The curves show essentially the same behavior as Fig.~\ref{e-hei-mu}, meaning that the model-related errors connected with eq.~(\ref{exc-c}) are almost always negligible compared to the discretization errors.

\section{Ultrashort time scales\label{fast-sect}}
In the previous sections the limits of the micromagnetic approximation have been discussed in the static case, except for a short remark on the D\"oring wall mass in section \ref{wall-sect}. But the micromagnetic approximations also reach their limit in the case of ultrashort time scales, especially in ultrafast laser-induced magnetization processes \cite{beaurepaire_ultrafast_1996,stanciu_all-optical_2007,vahaplar_ultrafast_2009}. The Landau-Lifshitz-Gilbert equation (\ref{LLG}) may then need to be generalized, which is usually done by replacing it with the Landau-Lifshitz-Bloch (LLB) equation \cite{garanin_fokker-planck_1997,kazantseva_towards_2008}. The LLB equation accounts for the temporal and spatial reduction of the magnitude of the magnetization, in contrast to the LLG equation that preserves the modulus $M_s=|\bm{M}|$. 
The microscopic details governing ultrafast laser-induced magnetization processes are still not completely understood, but it can be assumed that also in these cases at least temporarily very inhomogeneous structures can occur. Examining the range of validity of dynamic micromagnetism in such extreme situations is part of ongoing research, including some controversial aspects the discussion of which would go far beyond the scope of this article. For the sake of completeness it is however worth pointing out that static structures with strong inhomogeneities are not the only situation where micromagnetism can reach its limits. The common denominator between the strongly inhomogeneous structures discussed before and ultrafast magnetization processes is a high local energy density. Whenever a very high energy density develops (at least locally), the validity of the micromagnetic framework is challenged.

\section{Conclusions}
When highly inhomogeneous structures develop, the exchange length $l$ loses its meaning as a useful estimate for the upper bound of admissible cell sizes in  simulations. Strong inhomogeneities lead to numerical and systematic errors in the calculation of the exchange energy if standard micromagnetic theory is employed. While discretization errors systematically underestimate the exchange energy, methodological errors systematically overestimate it. Nevertheless, those two opposite effect generally do not compensate. When singularities occur in micromagnetic simulations, a careful analysis is required to ensure accurate results. Micromagnetic simulations involving singularities should be checked for an extrapolation to infinite discretization density in order to remove the discretization error. In order to obtain a more realistic description of Bloch points, the standard micromagnetic theory must be replaced or extended. A straightforward way of doing this is to abandon the assumption of constant magnitude of the magnetization by using the LLB equation also in the low-temperature range, as proposed recently by Lebecki {\em et al.} \cite{lebecki_key_2012}. With the LLB equation, a high energy exchange density can be used as an indicator for a locally reduced value of $M_s$, which can be corrected accordingly. This approach requires a careful calibration of the function $M_s=M_s(e_{\rm xc}^\mu)$ which enters as a degree of freedom in the LLB equation. A more accurate and parameter-free solution can be obtained by combining atomistic calculations which consider the Heisenberg interaction for a realistic atomistic lattice and to embed them into standard micromagnetic simulations \cite{hrkac_role_2010}. When the atomistically inhomogeneous magnetic region can be identified beforehand, {\em e.g.}, by the granular structure of the magnetic material \cite{kronmuller_micromagnetism_1997}, such a combined Heisenberg-continuum approach poses much less practical difficulties than an algorithm that detects and implements atomistic regions automatically. Such an approach is required in the case of a moving Bloch point. Developing a dynamic multimodel algorithm for such multiscale simulations represents a formidable programming endeavor.


%
\bibliographystyle{elsarticle-num}

\end{document}